\documentclass[twocolumn,prb,showpacs]{revtex4}

\usepackage{amssymb}
\usepackage{graphicx}
\usepackage{epsfig}
\usepackage{dcolumn}
\usepackage{bm}
\usepackage{amsmath}
\begin{document}
\title{Frequency doubling and memory effects in the Spin Hall Effect}
\author{Yuriy V. Pershin}
\email{pershin@physics.sc.edu} \affiliation{Department of Physics
and Astronomy and USC Nanocenter, University of South Carolina,
Columbia, SC, 29208}
\author{Massimiliano Di Ventra}
\email{diventra@physics.ucsd.edu} \affiliation{Department of
Physics, University of California, San Diego, La Jolla, California
92093-0319}

\begin{abstract}
We predict that when an alternating voltage is applied to a
semiconducting system with inhomogeneous electron density in the
direction perpendicular to main current flow, the spin Hall effect
results in a transverse voltage containing a double-frequency
component. We also demonstrate that there is a phase shift between
applied and transverse voltage oscillations, related to the
general memristive behavior of semiconductor spintronic systems. A
different method to achieve frequency doubling based on the
inverse spin Hall effect is also discussed.
\end{abstract}


\maketitle

In optics, frequency doubling (also called second harmonic
generation) is obtained from nonlinear processes, in which the
frequency of photons interacting with a nonlinear material is
doubled. This phenomenon was first observed in 1961 \cite{SG1} and
has found numerous applications in diverse areas of science and engineering
\cite{SG2}. Physically, the fundamental (pump)
wave propagating through a crystal with $\chi^{(2)}$
nonlinearity (due to the lack of inversion symmetry) generates a
nonlinear polarization which oscillates with twice the fundamental
frequency radiating an electromagnetic field with this doubled
frequency. In electronics, frequency doubling is a fundamental
operation for both analog and digital systems, which is however
achieved via complex circuits made of both passive and active
circuit elements \cite{freqd}.

As we demonstrate in this Letter, the possibility of generating
frequency doubling need not be limited to optical processes in
crystals, or require complex circuits. In fact, we show it can be
realized via a completely different physical mechanism using the
spin Hall effect \cite{spinHall}. Our idea is to use a material
with inhomogeneous doping in the direction perpendicular to the
main current flow. As it was recently shown~\cite{vprobe},
a dc voltage applied to such a system results in a transverse
voltage, similar to the Hall voltage, but with a different
symmetry: the sign of the transverse voltage due to the spin Hall
effect does not depend on the polarity of the applied field.
Therefore, when an ac voltage is applied, the transverse voltage
oscillations are similar in the positive and negative half-periods
of the applied voltage, which results in the frequency doubling.

\begin{figure}[b]
 \begin{center}
\includegraphics[angle=270,width=6.0cm]{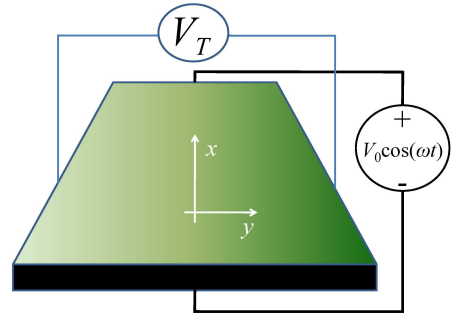}
\caption{\label{fig1}(Color online) Schematics of the experimental
setup. Inhomogeneous charge density is represented via a color
gradient.}
 \end{center}
\end{figure}

Moreover, the transverse voltage oscillations show hysteretic
behavior at different frequencies. This result is reminiscent of the recent experimental demonstration of
memory-resistive (memristive) behavior in certain nanoscale systems
\cite{memr1,memr2} and is consistent with our suggestion \cite{spinmemr} that some
semiconductor spintronic systems are intrinsically memristive
systems~\cite{chua76}. When a time-dependent voltage is applied to
such systems, their response is delayed because the adjustment
of spin polarization to changing driving field requires some time (due to
spin relaxation and diffusion processes). In other words, the
electron spin polarization has a short-time memory on its previous
state \cite{spinmemr}. A unique feature of the system
investigated in this work is that effects of spin memory manifest themselves in
the voltage response, while in the previous study \cite{spinmemr}
spin memory effects were predicted in the current. Below, we study
the frequency doubling and manifestation of spin memristive effects
both analytically and numerically. This work reveals fundamental aspects of the spin Hall effect that have not been explored yet, as well as its
possible use in electronic circuits.

Fig. \ref{fig1} shows a possible experimental setup which can be
used to observe frequency doubling by using the spin Hall effect.
An alternating voltage $V(t)=V_0\cos\left( \omega t \right)$ of frequency $\omega$ and
amplitude $V_0$ is
applied along a sample of semiconductor material (we call this $x$
direction). The electron density in the semiconductor is
inhomogeneous in the direction ($y$) perpendicular to main current
flow. A transverse time-dependent voltage $V_T$ then develops
between the sample boundaries $y=0$ and $y=L$, where $L$ is the
sample width. According to our previous calculations \cite{vprobe}
and recent experimental results \cite{miah}, it is expected that
the transverse voltage oscillation amplitude is a non-linear
function of $V_0$.

To study this system, we employ a self-consistent two-component drift-diffusion
model \cite{flatte,PDV} which is appropriate for the physics of this problem. The
inhomogeneous charge density profile $n(y)$ is defined via an
assigned positive background density profile $N(y)$ (such as the one
shown in Fig.~\ref{fig1}), which can be obtained in different ways
including inhomogeneous doping, variation of sample height or
gate-induced variation of electron density. Assuming homogeneous
charge and current densities in the $x$ direction and homogeneous
$x$-component of the electric field in both $x$ and $y$ directions,
the set of equations to be solved is

\begin{equation}
e\frac{\partial n_{\uparrow (\downarrow)}}{\partial
t}=\textnormal{div}  j_{y,\uparrow
(\downarrow)}+\frac{e}{2\tau_{sf}}\left(n_{\downarrow
(\uparrow)}-n_{\uparrow (\downarrow)} \right), \label{contEq}
\end{equation}
\begin{equation}
j_{y,\uparrow (\downarrow)}=\sigma_{\uparrow (\downarrow)}
E_y+eD\nabla  n_{\uparrow (\downarrow)}\pm \gamma I_{x,\uparrow
(\downarrow)} , \label{currentEq}
\end{equation}
and
\begin{equation}
\textnormal{div}E_y=\frac{e}{\varepsilon\varepsilon_0}\left(
N(y)-n\right), \label{puaeq}
\end{equation}
where $-e$ is the electron charge, $n_{\uparrow (\downarrow)}$ is
the density of spin-up (spin-down) electrons, $j_{y,\uparrow
(\downarrow)}$ is the current density, $\tau_{sf}$ is the spin
relaxation time, $\sigma_{\uparrow (\downarrow)}=en_{\uparrow
(\downarrow)}\mu$ is the spin-up (spin-down) conductivity, $\mu$
is the mobility, $D$ is the diffusion coefficient, $\epsilon$ is
the permittivity of the bulk, and $\gamma$ is the parameter
describing deflection of spin-up (+) and spin-down (-) electrons.
The current $I_{x,\uparrow (\downarrow)}$ in $x$-direction is
coupled to the homogeneous electric field $E\left(t
\right)=E_0\cos\left(\omega t\right)$ in the same direction as
$I_{x,\uparrow (\downarrow)}=en_{\uparrow (\downarrow)}\mu E\left(
t \right)$. The last term in Eq. (\ref{currentEq}) is responsible
for the spin Hall effect.

Equation (\ref{contEq}) is the continuity relation that takes into
account spin relaxation and Eq. (\ref{puaeq}) is the Poisson
equation. Equation (\ref{currentEq}) is the expression for the
current in $y$ direction which includes drift, diffusion and spin
Hall effect components. We assume here for simplicity that $D$,
$\mu$, $\tau_{sf}$ and $\gamma$ are equal for spin-up and spin-down
electrons.~\cite{prec} In our model, as it follows from Eq.
(\ref{currentEq}), the spin Hall correction to spin-up (spin-down)
current (the last term in Eq. \ref{currentEq}) is simply
proportional to the local spin-up (spin-down) density. All
information about the microscopic mechanisms for the spin Hall
effect is therefore lumped in the parameter $\gamma$.

Combining equations (\ref{contEq}) and (\ref{currentEq}) for
different spin components we can get the following equations for the
electron density $n=n_{\uparrow}+n_{\downarrow}$ and the spin
density imbalance $P=n_{\uparrow}-n_{\downarrow}$:
\begin{equation} \frac{\partial
n}{\partial t}=\frac{\partial}{\partial y} \left[ \mu n E_y + D
\frac{\partial n}{\partial y} +\gamma P \mu E\left( t \right)
\right] \label{CC}
\end{equation}
and
\begin{equation} \frac{\partial P}{\partial t}=\frac{\partial}{\partial y}
\left[ \mu P E_y + D \frac{\partial P}{\partial y} +\gamma n \mu
E\left( t \right) \right]-\frac{P}{\tau_{sf}}. \label{Peq}
\end{equation}

{\it Analytical solution --} Before solving Eqs.
(\ref{puaeq})-(\ref{Peq}) numerically, an instructive analytical
result can be obtained in the specific case of exponential doping
profile $N(y)=A\exp \left( \alpha y\right)$, with $\alpha$ a
positive constant. At small values of $E_0$, we search for a
solution in the form $n=n_0+\delta n$, $E_y=E_{y,0}+\delta E_y$,
$P=P_0+\delta P$. Setting $E_0=0$, the leading terms in the above
expansions can be easily obtained (see also Ref.
\onlinecite{vprobe}): $n_0=N(y)=Ae^{\alpha y}$, $E_y=-\frac{D\alpha}{\mu}$, $P_0=0$.
Next, using Eq. (\ref{Peq}) and neglecting the term $\sim \delta n
E_0$, we obtain
\begin{equation}
\delta P=\frac{e^{\alpha y}\gamma A\alpha \mu
E_0}{\frac{1}{\tau_{sf}^2} +\omega^2}\left[
\frac{1}{\tau_{sf}}\cos(\omega t)+ \omega \sin(\omega t) \right].
\label{Poft}
\end{equation}

Combining Eqs. (\ref{puaeq}),~(\ref{CC}), and~(\ref{Poft}),
integrating in $y$ and neglecting the term proportional to $\delta
n \delta E_y$ (this approximation neglects
small-amplitude higher harmonics terms), we obtain the following
equation for $\delta E_y$
\begin{eqnarray}
-\frac{\partial \delta E_y}{\partial t}= \frac{e \mu A e^{\alpha
y}}{\varepsilon\varepsilon_0} \delta E_y+\frac{\partial \delta
E_y}{\partial y} D\alpha -\frac{\partial^2 \delta E_y}{\partial
y^2} D + \nonumber \\ + \frac{e e^{\alpha y}\gamma^2 A\alpha \mu^2
E_0^2}{ \varepsilon\varepsilon_0 \left(\frac{1}{\tau_{sf}^2}
+\omega^2\right)}\left[ \frac{1}{\tau_{sf}}\frac{ \cos(2\omega
t)+1}{2}+ \omega \frac{\sin(2\omega t)}{2} \right]. \label{eq13}
\end{eqnarray}
Equation (\ref{eq13}) already demonstrates that the driving term for
$\delta E_y$ involves a doubled frequency. In order to further
proceed, let us consider a sample of a finite width $L$, in which
doping level variations are not dramatic. Then, in the first term in
the right hand side of Eq. (\ref{eq13}), we can write approximately
$e^{\alpha y}\simeq e^{\alpha y^*}$, where $0<y^*<L$. This
approximation allows us to find
\begin{equation}
\delta E_y=\left( C_1+\sqrt{C_2^2+C_3^2}\cos (2\omega t-\theta )
\right) e^{\alpha y},
\end{equation}
where $\theta$, defined as $\tan \theta=C_2/ C_3$, is a phase
shift, and
\begin{eqnarray}
C_1&=&-\frac{G}{2}\frac{1}{\tau_{sf} \mu A e^{\alpha y^*}}; \;\;\;G=\frac{\gamma^2 A\alpha \mu^2 E_0^2}{\frac{1}{\tau_{sf}^2}
+\omega^2},\\
C_2&=&-\frac{G}{2}\frac{\frac{1}{\tau_{sf}}+
\frac{e}{2\varepsilon\varepsilon_0}\mu A e^{\alpha
y^*}}{\frac{2\varepsilon\varepsilon_0\omega}{e}+\frac{e}{2\varepsilon\varepsilon_0\omega}\left(
\mu A e^{\alpha y^*} \right)^2 }, \\
C_3&=&\frac{G}{2}\frac{\omega-\frac{1}{\tau_{sf}}\frac{e}{2\varepsilon\varepsilon_0\omega}\mu
A e^{\alpha
y^*}}{\frac{2\varepsilon\varepsilon_0\omega}{e}+\frac{e}{2\varepsilon\varepsilon_0\omega}\left(
\mu A e^{\alpha y^*} \right)^2 }.
\end{eqnarray}

Finally, the transverse voltage is given by
\begin{equation}
\begin{split}
V_T&=-\int\limits_0^L \delta E_ydy \\
&= \left( C_1+\sqrt{C_2^2+C_3^2}\cos (2\omega t-\vartheta) \right)
\frac{1-e^{\alpha L}}{\alpha}.
\end{split}\label{deltaE2}
\end{equation}

Equation (\ref{deltaE2}) demonstrates that there are two
contributions to the transverse voltage $V_T$: a shift term
(proportional to $C_1$) and a double-frequency phase-shifted
oscillation term (proportional to $\sqrt{C_2^2+C_3^2}$). We have
found that Eq. (\ref{deltaE2}) is in excellent agreement with
results of our numerical calculations (given below) with the only
one adjustable parameter $y^*$. For a particular set of parameters
used below, a perfect match between analytical and numerical
calculations was obtained at $y^*=L/1.65$.

\begin{figure}[tb]
 \begin{center}
\includegraphics[,width=8.0cm]{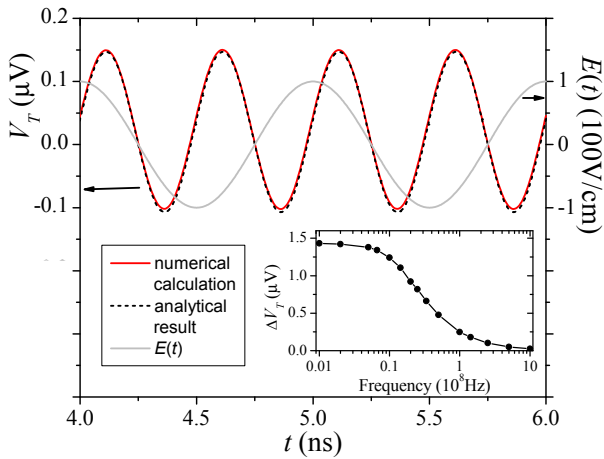}
\caption{\label{fig2}(Color online) Frequency doubling in
transverse voltage oscillations calculated at $\nu=10^8$Hz. A
phase shift between the applied field $E(t)$ solid line) and the
transverse field $V_T$ (red solid line) oscillations is clearly
seen. Inset: the transverse voltage oscillation amplitude as a
function of the applied voltage frequency (the dots represent
calculated values of $\Delta V_T$, the solid line is a fit to this
points). The plots were obtained using the parameter values
 $\mu=8500$cm$^2$/(Vs), $D=55$cm$^2$/s, $\varepsilon=12.4$,
 $\tau_{sf}=10$ns, $\gamma=10^{-3}$, $E_0=100$V/cm and the background density
 profile $N=10^{16}\textnormal{exp}(2y/L)$cm$^{-3}$, where $L=100\mu$m is the sample width and $0\leq y\leq L$.
The analytical curve (black dashed line) was obtained at
$y^*=L/1.65$.
 }
 \end{center}
\end{figure}

{\it Numerical solution --} Equations (\ref{puaeq})-(\ref{Peq}) can
be solved numerically for any reasonable form of $N(y)$. We choose
an exponential profile for its simplicity, the possibility to
realize it in practice, and for the purpose of comparison with the
above analytical results. We solve these equations iteratively,
starting with the electron density $n(y)$ close to $N(y)$ and $P(y)$
close to zero and recalculating $E_y(y)$ at each time
step.~\cite{prec1} At each time step, the transverse voltage is
calculated as a change of the electrostatic potential across the
sample.

\begin{figure}[tb]
 \begin{center}
\includegraphics[width=8.0cm]{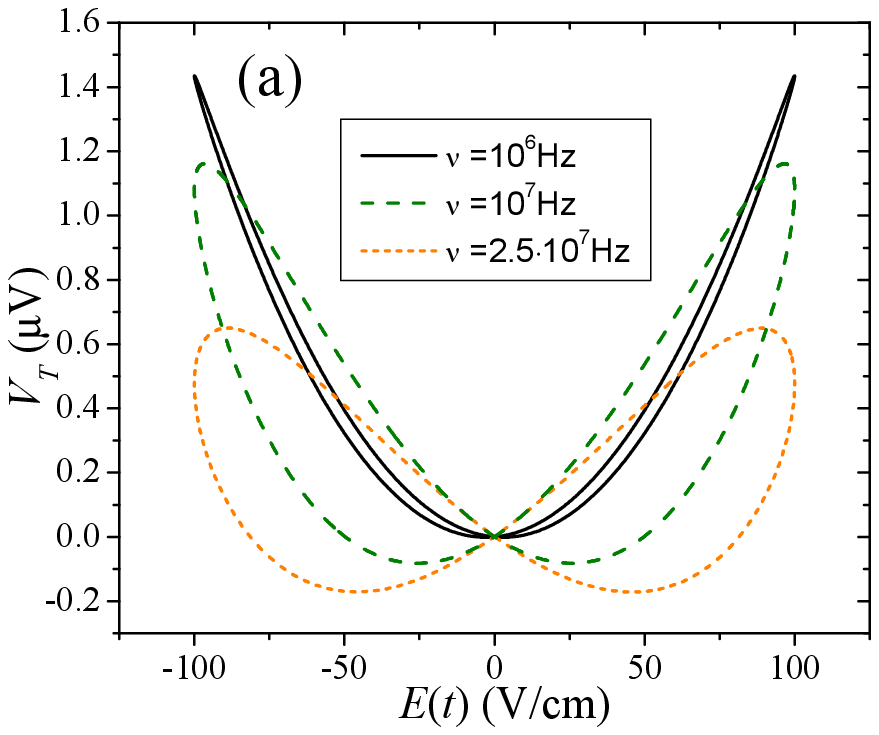}
\includegraphics[width=8.0cm]{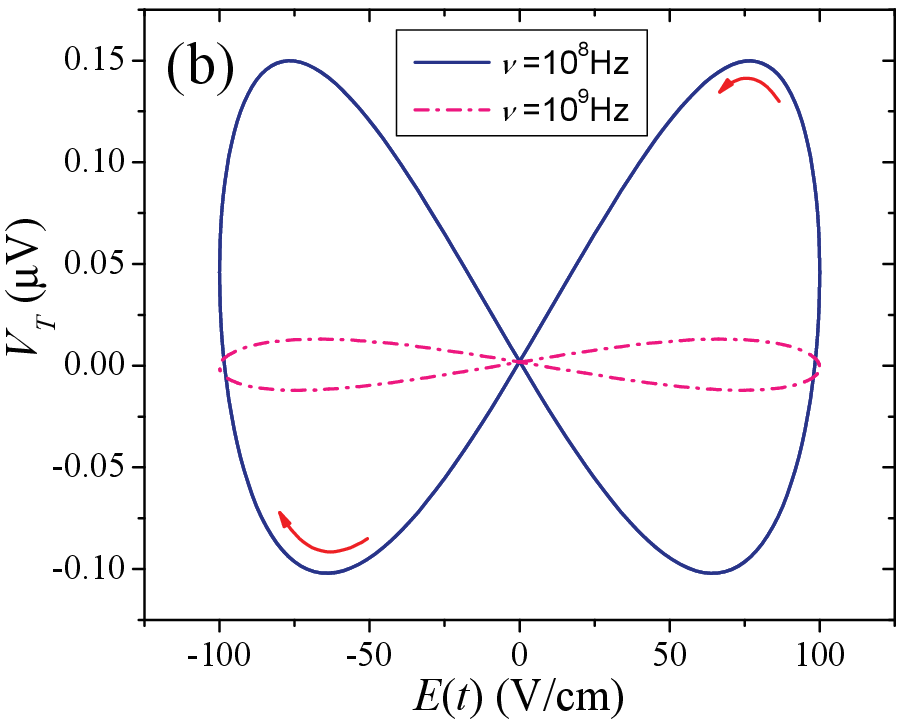}
\caption{\label{fig3}(Color online) The transverse voltage as a
function of applied electric field of (a) lower and (b) higher
frequencies.}
 \end{center}
\end{figure}

Selected results of our numerical calculations are presented in
Figs. \ref{fig2} and \ref{fig3}. In particular, Fig. \ref{fig2}
demonstrates that the transverse voltage oscillations are indeed
of a doubled-frequency character and phase-shifted with respect to
the applied voltage. Another important feature shown in Fig.
\ref{fig2} is the excellent agreement between our analytical and
numerical results. This agreement was obtained by an appropriate
choice of the parameter $y^*$ defined after Eq. (\ref{eq13}). We
observed that $y^*$ slightly depends on $E_0$ and $\omega$, and,
once $y^*$ is selected, the numerical and analytical solutions are
in a good agreement in a wide range of excitation voltage
parameters.

Multiple signatures of spin memristive behavior are clearly seen in
Figs. \ref{fig2} and \ref{fig3} including, constant and phase shifts
in $V_T$ depicted in Fig. \ref{fig2}, frequency dependence of the
transverse voltage oscillation amplitude shown in the inset of Fig.
\ref{fig2}, and hysteresis behavior plotted in Fig. \ref{fig3}. All
these features have a common origin: the adjustment of electron spin
polarization to changing voltage takes some time. In particular, at
low frequencies, we observe a small hysteresis in Fig. \ref{fig3}(a)
because when the applied electric field is changed slowly (on $V_T$
equilibration time scale), at each moment of time $t$ the
instantaneous $V_T$ is very close to its equilibrium value
irrespective of the driving field $E(t)$. At high frequencies, the
situation is opposite: when the applied electric field changes very
fast, the electrons ``experience'' an average (close to zero)
applied electric field, resulting in a significantly reduced
transverse voltage oscillations amplitude.

We also note that at those moments of time when $E(t)=0$, the
transverse voltage is very close to, but not exactly, zero. This
small deviation from an ideal memristive behavior~\cite{chua76}
(predicting $V_T=0$ when $E(t)=0$ and related to the absence of
energy storage) should be a common feature of solid-state
memristive systems operating at frequencies comparable to the inverse characteristic time of
charge equilibration processes.

\begin{figure}[tb]
 \begin{center}
\includegraphics[width=6.0cm]{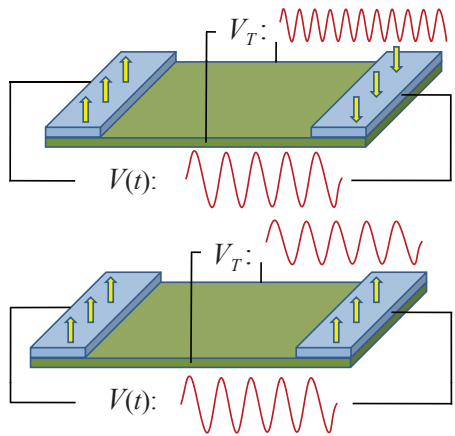}
\caption{\label{fig4}(Color online) Frequency doubling with the
inverse spin Hall effect where a nonmagnetic material (green) is
sandwiched between two ferromagnetic leads (blue). The transverse
voltage oscillation frequency depends on the relative orientation of
the magnetization of the ferromagnetic contacts. The frequency is
doubled when the direction of magnetization of ferromagnetic
contacts is antiparallel. When the direction of magnetization of the
ferromagnetic contacts is parallel, the transverse voltage frequency
is equal to the applied voltage frequency. The oscillation
amplitudes are not to scale.}
 \end{center}
\end{figure}

We conclude by noting that another way to realize the frequency
doubling discussed in this paper can be realized by sandwiching a
nonmagnetic homogeneous material between two ferromagnets (see
schematic in Fig. \ref{fig4}). In this case one can employ the {\em
inverse} spin Hall effect, in which the spin current flowing in the
nonmagnetic material induces transverse charge current, and thus
causes charge accumulation \cite{ishe1,ishe2,ishe3,ishe4,ISH}. Since
the electromotive force in the inverse spin Hall effect is
\cite{ISH}
 $\sim \vec
J_S\times\vec\sigma$, where $\vec J_S$ is the spin current along
the sample and $\vec\sigma$ is the Pauli matrix, the simultaneous
change of the current direction and its spin polarization, that
occurs at antiparallel magnetization of ferromagnetic contacts,
does not change the electromotive force polarity leading to
transverse voltage oscillations with a doubled frequency. On the
other hand, for parallel orientation of the spin polarization of
the ferromagnetic contacts, the direction of spin current changes
within each voltage cycle but not the the direction of spin
polarization. This thus results in a transverse voltage with the
same frequency as the longitudinal one, albeit with lower
amplitude (see Fig. \ref{fig4}).  However, since spin injection
allows for much higher levels of electron spin polarization (of
the order of several tens of percents) compared to the spin Hall
effect (normally, less than one percent), we expect the amplitude
of $V_T$ oscillations to be larger in the inverse spin Hall effect.
An external magnetic field can also be used in such experiments as
an additional control parameter.

Finally, the phenomena we predict can be easily verified
experimentally. They provide additional insight on the spin Hall
effect and may find application in, e.g., analog electronics
requiring frequency doubling. We thus hope our work will motivate
experiments in this direction.

This work has been partially funded by the NSF grant No.
DMR-0802830.


\begin{thebibliography}{}
\bibitem{prec} This is a good approximation for the range of
parameters considered in this work.
\end{thebibliography}
\end{document}